\documentclass[epj]{svjour}
\usepackage{graphicx}
\usepackage{bbm}

\newcommand {\apgt} {\ {\raise-.5ex\hbox{$\buildrel>\over\sim$}}\ }
\newcommand {\aplt} {\ {\raise-.5ex\hbox{$\buildrel<\over\sim$}}\ } 

\begin{document}
\title{Boltzmann-Gibbs thermal equilibrium distribution for classical systems 
  and Newton law: A computational discussion}  
\author{Fulvio Baldovin\inst{1}, 
  Luis G. Moyano\inst{2,}\inst{3} and 
  Constantino Tsallis\inst{2,}\inst{3}}
%\email{fulvio.baldovin@pd.infn.it}
%\offprints{}
\institute{
1 {\it Dipartimento di Fisica and
Sezione INFN, Universit\`a di Padova,\\
\it Via Marzolo 8, I-35131 Padova, Italy}\\ \email{baldovin@pd.infn.it}\\
2 {\it Centro Brasileiro de Pesquisas F\'{\i}sicas \\
Rua Xavier Sigaud 150, 22290-180 Rio de Janeiro-RJ, Brazil}\\ \email{moyano@cbpf.br, tsallis@santafe.edu}\\
3 {\it Santa Fe Institute,\\
1399 Hyde Park Road, Santa Fe, New Mexico 87501, USA}
}
\titlerunning{Boltzmann-Gibbs thermal equilibrium distribution and Newton law}
\authorrunning{Baldovin et al}
%\date{Received: date / Revised version: date}

%\author{Luis G. Moyano and Constantino Tsallis}
%\email{moyano@cbpf.br, tsallis@santafe.edu}
%\affiliation{
%Centro Brasileiro de Pesquisas F\'{\i}sicas\\
%Rua Xavier Sigaud 150, 
%22290-180 Rio de Janeiro-RJ, Brazil
%}
%\affiliation{
%Santa Fe Institute,
%1399 Hyde Park Road,
%Santa Fe, New Mexico 87501, USA\\and
%}
\abstract{
We implement a general numerical calculation that allows for a direct
comparison between nonlinear Hamiltonian dynamics and the
Boltzmann-Gibbs canonical distribution in Gibbs $\Gamma$-space.
Using paradigmatic first-neighbor models, namely, the inertial
$XY$ ferromagnet and the Fermi-Pasta-Ulam 
$\beta$-model, 
we show that at intermediate energies the Boltzmann-Gibbs equilibrium
distribution is a consequence of Newton
second law (${\mathbf F}=m{\mathbf a}$).  
At higher energies we discuss partial agreement between time and
ensemble averages.  
}
\maketitle

%PACS numbers: 05.10.-a, 05.20.-y, 05.45.-a, 05.20.Gg 

%\begin{multicols}{2}

\begin{sloppypar}
The problem of the dynamical foundation of Boltzmann-Gibbs 
(BG) statistical
mechanics dates back to the original proposal of this
powerful formalism (see, e.g., \cite{einstein_01})
and despite many important results this
fundamental question \cite{huang_01} still
presents open basic aspects
(see, e.g., \cite{takens_01,carati_01,livi_01,cohen_01,casetti_01} and
references therein). 
Thanks to the current computational capability we can numerically
integrate the Hamilton equations of large enough systems
and compare the
results with the predictions of the BG formalism.
Although this technique has been largely and successfully implemented 
in a microcanonical perspective (fixed-energy molecular
dynamics), the methods used when addressing systems in contact with a
thermostat
(such as Monte Carlo
and Nos\'e-Hoover \cite{frenkel})
usually impose an \mbox{{\it ad hoc}} dynamics.
In this paper we introduce a scheme which 
enables the discussion of the canonical distribution 
in Gibbs \mbox{$\Gamma$-space} on the basis
of the equations of motions.
Within the present approach both time and ensemble averages
are performed dynamically, so that we are able to discuss 
ergodicity.
Using two paradigmatic first-neighbor nonlinear Hamiltonian
systems --- the one-dimensional inertial 
$XY$ ferromagnet and the Fermi-Pasta-Ulam (FPU) $\beta$-model  ---  
we find a remarkable agreement between BG equilibrium calculations and
dynamical ensemble averages. We also compare partial ergodicity
failure with the maximum Lyapunov coefficient.
Our numerical calculation can be implemented in systems
that allow for a textbook definition of
canonical ensemble (i.e., part of a large isolated system).
It would also be interesting to check the same procedure in
situations where, due for example to the presence of
long-range terms, important deviations from the BG predictions have
been found \cite{latora_01,gell_mann_01}. 
We are presently making progress on this task.
\end{sloppypar}

Given some macroscopic conditions 
in the phase space of
the system under consideration ($\Gamma$-space), 
the average value of a dynamical function can be defined
using time or ensemble averages; 
ergodicity means that these two methods are 
equivalent. 
We remark that both approaches are dynamically realizable.
In the first case one 
focuses on a single dynamical realization.
The probability $p_R$ of finding the system inside a
coarse-grained region $R$ of \mbox{$\Gamma$-space}
is defined by the fraction of time $t_R$ spent by the system
inside that region during the (eventually infinite) 
total amount of time $\tau$ of its phase space trajectory:
$p_R^{\;t}\equiv t_R/\tau$, where the superscript $t$ stands
for {\it time} definition.
The second is achieved for instance by fixing 
a certain instant of time $t^\ast$ and
repeating the dynamical evolution up to $t^\ast$, under
the same macroscopic (but different microscopic) initial conditions. 
Counting the number of
times $n_R$ the system is found in region $R$ at time
$t^\ast$, 
with respect to the (eventually infinite) total number of
times $n$ the calculation is performed, one defines
$p_R^{\;e}\equiv n_R/n$, where the superscript $e$ indicates
{\it ensemble}.

For a typical  $N$-body conservative Hamiltonian system
 (typical in the sense that it complies with the BG
prescriptions)
at fixed energy $E_N$ (microcanonical setup), 
a standard introduction of the canonical ensemble 
is obtained defining the canonical system as
composed by a subset of $M$ interacting elements, 
with $1\ll M\ll N$. 
The energy of the $M$ elements satisfies
$E_M \ll E_N$, and the interaction energy between the canonical 
system and the rest of the isolated system  (thermal bath) is assumed to be much smaller
than $E_M$.
Under these circumstances, the probability $p_j$ of finding the
system in a $M$-microstate $j$ is given by the 
BG equilibrium calculation
$p_j\propto e^{-\beta E_j}$, where $\beta\equiv 1/T$ is the
inverse temperature 
(without loss
of generality, we set the Boltzmann constant $k_B\equiv 1$), 
and $E_j$ is the energy of the microstate. 
A dynamical approach for the confirmation of this result
must face the
following numerical difficulty.
The $\Gamma$-space is $Md$ dimensional, $d$ being the
dimension of the single-particle phase space.
If we implement a coarse-graining for example by making  
a partition of $k$ intervals in each coordinate, the
total number of (hyper)cells $\Omega_M$ is of order
$k^{Md}$. 
Just to put some indicative numbers, 
with $k=4$, $M=100$ and $d=2$ we get 
$\Omega_M\sim 4^{200}\sim 10^{120}$.
We should hence implement a numerical integration of 
$2N(\gg200)$ 
Hamilton equations with a total amount of time $\tau$ (or
a total number of realizations $n$) much larger than
$10^{120}$, which is beyond what we can presently do
numerically. 

\begin{sloppypar}
Nevertheless, we can proceed through an alternative
path and, instead of focusing on the probability associated to a microstate, we
could consider the probability of finding the
canonical system 
with a given energy $E_M$. In this case the BG answer is 
\begin{equation}
p(E_M)=\frac{\omega(E_M) e^{-\beta E_M}}{Z},
\end{equation}
where $Z$ is the partition function and 
\begin{equation}
\omega(E_M)=\int\prod_{i=1}^M(d p_id q_i)\delta[E_M-H_M(p_i,q_i)]
\end{equation}
is the phase-space density of states at
energy $E_M$. As well known for a classical system,  $\omega(E_M)$ does not depend 
on any particular thermal statistics, it only depends on the Hamiltonian
of the system. In other words, we can calculate $\omega(E_M)$ by 
using {\it any} statistics, for example the BG one \cite{murilinho}. The density of states 
$\omega(E_M)$ can be analytically estimated through 
the thermodynamic relation linking the statistical entropy
to the temperature: $\partial\ln\omega(E)/\partial E=\beta$.
Integrating this relation we have that $\omega(E_M)$ is
given through the caloric curve $T(E)$:
\end{sloppypar}
\begin{equation}
\frac{\omega(E_M)}{\omega(E_0)}=
\exp\left[\int_{E_0}^{E_M} d E'\;\beta(E')\right],
\label{states_density}
\end{equation}
where $E_0$ is the energy of the fundamental state.
In brief, the Hamiltonian structure of the system defines
the density of states as a function of the energy; once this
relation is known it is sufficient to multiply $\omega(E_M)$
by the Boltzmann factor $e^{-\beta E_M}$ and to normalize, in
order to obtain $p(E_M)$ for the whole spectrum of
temperatures. 
Now, the dynamical computation of $p(E_M)$ is much easier than
the one for $p_j$. 
All we have to do is to numerically
integrate Hamilton equations and to calculate the value of
the energy $E_M$ for the canonical subset 
at each integration step. 
We can then coarse-grain the energy spectrum into bins
of width $\Delta E_M$
and build up a normalized histogram of the occurrence
of each of these bins. 
In analogy with the previous discussion, 
\begin{equation}
p^{\;t}(E_M)\equiv \frac{t(E_M)}{\tau\;\Delta E_M}
\;\;{\rm and}\;\;
p^{\;e}(E_M)\equiv \frac{n(E_M)}{n\;\Delta E_M}
\end{equation}
represent then the probability distribution of finding the
canonical system with energy $E_M$, respectively using
time and ensemble averages.

\begin{figure}
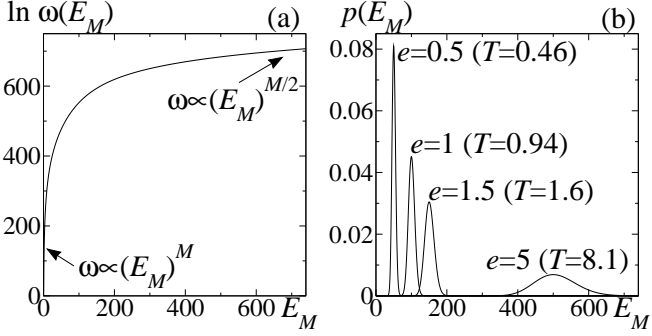

\begin{center}
\includegraphics[width=0.49\columnwidth,angle=0]{fig_1_a.eps}
\includegraphics[width=0.49\columnwidth,angle=0]{fig_1_b.eps}
\end{center}
\caption{\small 
BG analytical canonical prediction for the inertial
$XY$-ferromagnetic-rotors with $M=100$.   
(a) Logarithm of the density of states $\omega_M(E_M)$.
(b) $p(E_M)\equiv\omega_M(E_M)\exp(-E_M/T)/Z$, for different
temperatures.   
} 
\label{fig_density}
\end{figure}

\begin{sloppypar}
To illustrate this calculation, we consider next 
a specific class of analytically solvable nonlinear 
first-neighbor Hamiltonians, 
\begin{equation}
H_N=K_N+V_N=\sum_{i=1}^N\left[\frac{p_i^2}{2}+V(q_{i+1}-q_i)\right],
\end{equation}
with periodic boundary conditions ($q_{N+1}\equiv q_1$).  
As a first case we analyze a one-dimensional chain of
rotors with $V(q_{i+1}-q_i)\equiv 1-\cos(q_{i+1}-q_i)$, 
so that the canonical coordinates $q_i\in[0,2\pi)$
and $p_i\in{\mathbbm R}$ are respectively the angular
coordinates and the angular momenta of the (unit inertial
momenta) rotors. 
This Hamiltonian is an inertial version of the classical $XY$
ferromagnetic spin model and constitutes a dynamical
prototype for spin systems in statistical mechanics \cite{livi_01,casetti_01}.  
The model is nearly integrable for both low and high
energies. The former regime is defined for $T<0.05$
(specific energy $e<0.05$) \cite{livi_01} and it is called 
strong coupling regime, for which the rotors constitute a
set of oscillators almost linearly coupled. The latter is
obtained say for $T>10$ ($e>6$) \cite{livi_01}, where the
rotors are almost free (weak coupling regime).
For this model, dynamical deviation from BG statistics where detected
both in the strong and in the weak coupling regimes. Since our main
scope is to check our calculation scheme in standard situations, we
will mainly concentrate in the intermediate energy range, and discuss
partial disagreement that occurs at higher energies. 
The canonical partition function 
\begin{equation}
Z_M=\int\prod_{i=1}^M (d p_id q_i)\exp\left[-\beta
H_M(p_i,q_i)\right],
\end{equation}
gives, for this model, 
the specific free energy 
\mbox{$f \equiv -\lim_{M \to \infty}[\ln Z_M/(M\beta)]$}
(see. e.g., \cite{livi_01}):  
\begin{equation}
f=- T\left[
\frac{1}{2}\ln T+\ln I_0(\frac{1}{T})+\ln 2\pi^{\frac{3}{2}} 
\right]+1 \, ,
\label{free_energy_rotors}
\end{equation}
where
$I_0(x)$ is the modified Bessel function of the first
kind of order zero.
Inversion of the relation 
\mbox{$E(T)=F-T\partial F/\partial T$}
furnishes the BG caloric curve
$T(e)$, where \mbox{$e\equiv\lim_{M\to\infty}E_M/M$}.  
We then rescale the $e$-axis by a factor $M$ and use the fact that the
temperature  is an intensive parameter to get $T(E_M)$. 
The integration in Eq. (\ref{states_density}) finally gives 
$\omega(E_M)$ for any large-but-finite value of $M$.
In Fig. \ref{fig_density}(a) we plot the logarithm
of $\omega(E_M)$ for the first-neighbor rotors with $M=100$ and  
Fig. \ref{fig_density}(b) displays BG $p(E_M)$ 
for different values of the
temperature $T$ (or of the specific energy $e$).
We remark that, thanks to the elementary properties of the
logarithmic function, it is possible to
implement this calculation
for quite large values of $M$, since one essentially deals
with the exponents. 
\end{sloppypar}

\begin{sloppypar}
Because we are interested in very large values of $\tau$ and $n$,
the dynamical integration of Hamilton equations has been
performed using  
the $4$th order symplectic Neri-Yoshida integrator \cite{yoshida_01} 
with an iteration parameter that assures an
energy conservation 
$\Delta E_N /E_N \simeq10^{-3}$ (a few runs with $10^{-5}$ showed that
$10^{-3}$ is enough for our scopes).
In particular, we checked that the energy fluctuations of the total
system introduced by the finite precision of the integrator algorithm  
is order of magnitudes smaller than those that one would have in
presence of a thermal coupling.
An important point to perform an
efficient calculation concerns the initial
conditions, that must be close enough to equilibrium 
to avoid long transients. In this way we focus only on the equilibrium 
properties of the model, ruling out the possible presence of
metastable or quasi-stationary states 
appearing with far-from-equilibrium initial conditions. 
Since the system does not display any phase transition for
$T>0$ but presents a tendency to clusterization at low
temperatures, we have used a Maxwellian distribution for
the angular momenta (with the appropriate temperature) and
a set of $l$ equidistant Gaussian distributions for
the angles, each with the same variance appropriately
calculated in order to yield the desired total energy $E_N$. 
In our calculation it was sufficient to use $l=6$ for 
a fast enough relaxation to equilibrium in all our
(microcanonical) setups.  We have checked that this particular choice has no 
influence on the functional form of the equilibrium probability density functions: 
it is done to save computational time. Different close-to-equilibrium 
initial conditions 
eventually yield the same results. For all our results 
we have waited for
$10^3$ iteration steps before starting the measurements
for a canonical system which is composed by a
randomly chosen subset of $M$ adjacent rotors.
\end{sloppypar}

\begin{figure}
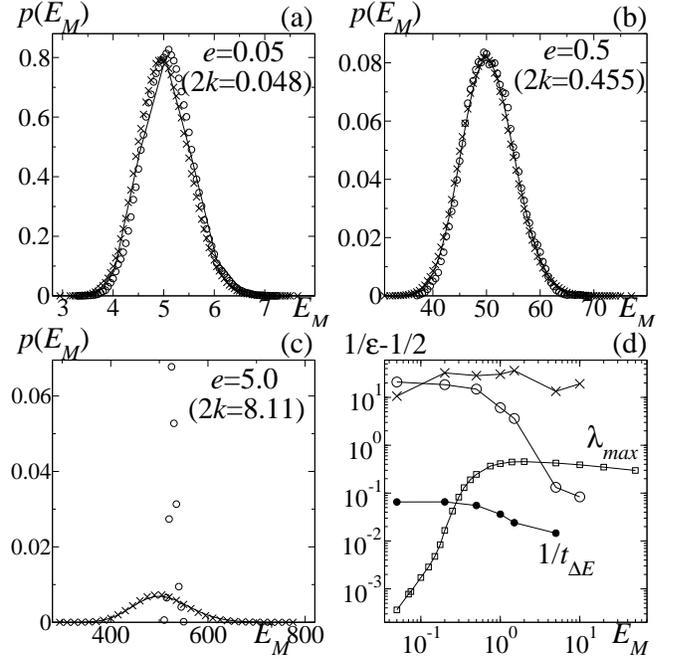

\begin{center}
\includegraphics[width=0.49\columnwidth,angle=0]{u0-05.eps}
\includegraphics[width=0.49\columnwidth,angle=0]{u0-5.eps}
%\\
%\vspace{2.5cm}
%\includegraphics[width=0.49\columnwidth,angle=0]{u1-5.eps}
\includegraphics[width=0.49\columnwidth,angle=0]{u5.eps}
\includegraphics[width=0.49\columnwidth,angle=0]{discrepancy.eps}
\end{center}
\caption{\small 
(a-c) Comparison between the BG prediction $p(E_M)$ (full
line), the ensemble dynamical average $p^{\;e}(E_M)$
(crosses), and the time dynamical average $p^{\;t}(E_M)$
(circles). 
$k\equiv K_M/M$ is the value of the specific kinetic energy.
(d) Analysis of discrepancy 
between $p^{\;e}(E_M)$ and $p(E_M)$ (crosses),
and $p^{\;t}(E_M)$ and $p(E_M)$ (empty circles).
We also plot the largest Lyapunov coefficient
$\lambda_{max}$ (squares) 
and the inverse of the time-scale for a normal fluctuation
$1/t_{\Delta E_M}$ (full circles). 
The lines are guides to the eye.
See text for details. 
} 
\label{fig_rotors}
\end{figure}

In Fig. \ref{fig_rotors}(a-c) we present a striking agreement
between the BG analytical prediction for $p(E_M)$ (full line)
and the dynamical estimation of $p^{\;e}(E_M)$ (crosses) for
various order of magnitudes of the specific energy $e$ 
with a setup $(M,N)=(10^2,10^3)$ and a total number of realizations
$n=5\times10^6$. On the other hand, $p^{\;t}(E_M)$ (circles), calculated with
a total number of iteration steps $\tau=5\times10^7$, displays a
good agreement with respect to the BG analytical
distribution for intermediate energies,
but starts to show large discrepancies when entering in the
weak-coupling regime.
In order to quantify this difference, we have defined the
discrepancy $0\leq\epsilon\leq2$ between two distributions as the
integral of the absolute value of the difference of the
distributions. 
To allow for a comparison with the largest Lyapunov
coefficient, in Fig. \ref{fig_rotors}(d) we plot the
quantity $0\leq1/\epsilon-1/2\leq\infty$ which is zero for maximum
discrepancy and infinite for perfect overlap of the distributions. 
While for ensemble averages $1/\epsilon-1/2$ is large and almost constant
with the energy, in the case of time averages 
such a quantity presents a dramatic decrease for large energies. 
In fact, we verified that the time necessary to have a typical energy
fluctuation of the canonical subset ($\Delta E_M\sim E_M/\sqrt M$)
grows with the energy (see full circles in Fig. \ref{fig_rotors}(d),
where we plot the inverse of this time),
as a consequence of the fact
that rotors are increasingly free (the potential
is upper bounded). 
We point out that the largest Lyapunov coefficient (squares in
Fig. \ref{fig_rotors}(d)) does not display a significant correlation
with the time characterizing
relaxation of $p^{\;t}(E_M)$ (circles in
Fig. \ref{fig_rotors}(a-c))  towards the BG distribution
$p(E_M)$ (see also \cite{livi_01} 
for a discussion of this point). 
This means that for the present system the positivity of the largest
Lyapunov coefficient is a measure of local chaos that does not imply
relaxation to global chaos. 

\begin{figure}
\begin{center}
\includegraphics[width=0.98\columnwidth,angle=0]{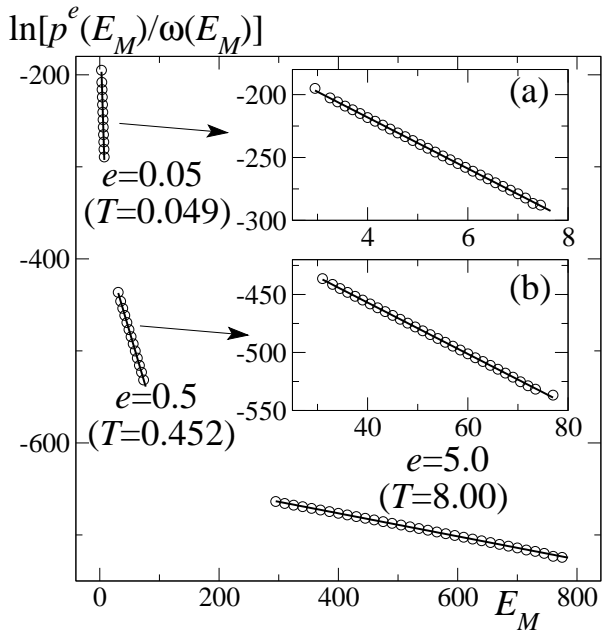}
\end{center}
\caption{\small 
Dynamical evidence of the Boltzmann factor. 
We plot $\ln[p^{\;e}(E_M)/\omega(E_M)]$
for the ensemble averages of Fig. \ref{fig_rotors} (circles).
$T$ is the reciprocal of the slope of linear regressions (full
lines) on the data. 
Insets (a) and (b) show a magnification of the results for
$e=0.05$ and $e=0.5$ respectively.
} 
\label{fig_boltzmann_factor}
\end{figure}

An important result is the coincidence between the value of
the BG temperature $T$ and twice the specific kinetic energy
$k\equiv K_M/M$ within an error of at most $2\%$. We stress that the 
probability density functions shown in Fig. \ref{fig_rotors} are 
obtained by means of first principles only and with complete 
independence from the BG theory, which we are checking. 
The concurrence between dynamics and the Boltzmann factor
appears satisfactorily in the linear regressions of
Fig. \ref{fig_boltzmann_factor}, 
where we plot $\ln[p^{\;e}(E_M)/\omega(E_M)]$ for the ensemble
averages of Fig. \ref{fig_rotors}(a-c).   
With other values of
$(M,N)$, namely $(50,500)$ and $(10^3,10^4)$, the results were 
qualitatively the same. 

\begin{figure}
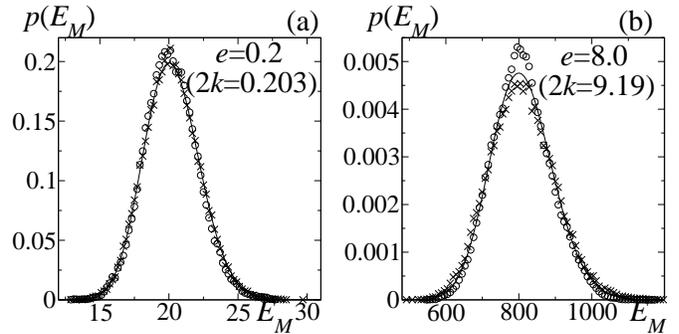

\begin{center}
\includegraphics[width=0.49\columnwidth,angle=0]{u0-2_fpu.eps}
\includegraphics[width=0.49\columnwidth,angle=0]{u8-0_fpu.eps}
\end{center}
\caption{\small 
Same as Fig. \ref{fig_rotors}(a-c) for the FPU $\beta$-model.
See text for details.
} 
\label{fig_fpu}
\end{figure}

We also obtained a confirmation of our results by
implementing the same calculation scheme for the FPU
$\beta$-model, defined by the potential 
$V(q_{i+1}-q_i)\equiv(q_{i+1}-q_i)^2/2+
0.1(q_{i+1}-q_i)^4/4$ with $q_i\in\mathbbm R$, 
again considering close-to-equilibrium initial conditions 
(see, e.g., \cite{livi_01} for the analytical canonical
solution and for a discussion of initial 
conditions). 
Although it is known that the FPU model presents, in common with the
rotors model, a very rich anomalous
behavior at low energies \cite{carati_01,livi_01}, for our initial
conditions and for the
energy-range we tested we found that $p^{\;t}(E_M)$ is in good agreement
with the BG prediction (Fig. \ref{fig_fpu}).

In summary, we recall that
using the standard BG formalism and common numerical
techniques, we have introduced a new calculation that allows
for a comparison between nonlinear Newtonian dynamics and
canonical statistical mechanics.  
Implementing a standard
setup we have in fact shown that the BG
energy distribution in $\Gamma$-space coincides with the one
that is obtained dynamically (integrating Hamilton
equations for close-to-equilibrium initial conditions) 
when an ensemble average is executed. 
We have checked this conclusion for two paradigmatic
first-neighbor nonlinear Hamiltonians.
As a side result, this calculation provides a dynamical
confirmation of the very well known relation between
temperature and specific kinetic energy $k=T/2$ 
(for one-dimensional systems).
With respect to finite-time
dynamical averages, at moderate low energies 
we have found a confirmation of the BG predictions.
For the $XY$-model at high energies, if the time-scale is not
very large, finite-time averages disagree
with ensemble averages as a consequence of an increase of
the time-scale of typical energy fluctuations. 
The energy dependence of this discrepancy does not display correlation with that       
of the largest Lyapunov coefficient (see also \cite{livi_01}).

Finally, let us emphasize that what we have shown here is
that equilibrium  thermal statistics descends
from (finite-precision) mechanics, 
even for a system in contact with a
thermostat (usually discussed through Monte Carlo
or Nos\'e-Hoover techniques, which do not deduce the equilibrium distribution
but impose it \cite{frenkel}). Indeed, this is the significance of
Figs. \ref{fig_rotors}(a-c) and \ref{fig_fpu}, where 
circles and crosses have been obtained from Newton law,
whereas full lines come from the BG
theory. Equivalently, if we recall that the density of
states is a purely mechanical concept, the same
conclusion is shown in Fig. \ref{fig_boltzmann_factor}.  
The present calculation scheme 
provides an insight onto the basic question
of the dynamical foundation of statistical mechanics
\cite{einstein_01,huang_01,livi_01,cohen_01,casetti_01}, 
and may serve as a useful tool in the discussion of complex
situations (see e.g., \cite{latora_01}) where dynamical
discrepancies with the BG theory have been found.

\section*{Acknowledgments} 
We thank useful remarks from L. Galgani, A. Rapisarda and S. Ruffo 
as well as partial support from
CAPES, PRONEX, CNPq and 
FAPERJ (Brazilian agencies).

%\end{multicols}

%\twocolumn


\begin{thebibliography}{99}

\bibitem{einstein_01} A. Einstein, Annalen der Physik {\bf
33}, 1275 (1910) [ 
``Usually $W$ is put equal to the number
of complexions... In order to calculate $W$, one needs a
{\it complete} (molecular-mechanical) theory of the system
under consideration. Therefore it is dubious whether the
Boltzmann principle has any meaning without a complete
molecular-mechanical theory or some other theory which
describes the elementary processes.  $S=\frac{R}{\cal N}\log
W+\;{\rm const}.$ seems without content, from a
phenomenological point of view, without giving in addition
such an {\it Elementartheorie}.'' (Translation from Abraham
Pais, {\it Subtle is the Lord...}, Oxford University Press,
1982) 
].

\bibitem{huang_01}
K. Huang, {\it Statistical Mechanics} (J. Wiley and
Sons, New York, 1987), p. 90-91 [
``We mentioned the ergodic theorem in Section 3.4, but did
  not use it as a basis for the microcanonical ensemble,
  even though, on the surface, it seems to be the
  justification we need. The reason is that existing proofs
  of the theorem all share (...) an avoidance of dynamics. For
  this reason, they cannot provide the true relaxation time
  for a system to reach local equilibrium (typically about
  $10^{-15}$ s for real systems), but have a
  characteristic time scale of the order of the Poincar\'e
  cycle. For this reason, the ergodic theorem has so far
  been an interesting mathematical exercise irrelevant to
  physics.'' ].   

\bibitem{takens_01}
F. Takens, in {\it Structures in dynamics - Finite 
dimensional deterministic studies}, eds. H.W. Broer, F. Dumortier, S.J. 
van Strien and F. Takens (North-Holland, Amsterdam, 1991), page 253 [``The values 
of $p_i$ are determined by the following 
dogma: if the energy of the system in the $i^{th}$ state is $E_i$ and 
if the temperature of the system is $T$ then: 
\mbox{$p_i=\exp\{-E_i/kT\}/Z(T)$}, where $Z(T)=\sum_i \exp\{-E_i/kT\}$, (this 
last constant is taken so that $\sum_i p_i=1$). This choice of $p_i$ is called {\it Gibbs distribution}. We shall 
give no justification for this dogma; even a physicist like Ruelle 
disposes of this question as ``deep and incompletely clarified". "]

\bibitem{carati_01}
A. Carati, L. Galgani and B. Pozzi, Phys. Rev. Lett. {\bf 90}, 010601
(2004); 
M.C. Carotta, C. Ferrario, G. Lo Vecchio, and L. Galgani,
Phys. Rev. A {\bf 17}, 786 (1978). 

\bibitem{livi_01}
R. Livi, M. Pettini, S. Ruffo and A. Vulpiani,
J. Stat. Phys. {\bf 48}, 539 (1987); see also 
D. Escande, H. Kantz, R. Livi and S. Ruffo, 
J. Stat. Phys. {\bf 76}, 605 (1994) .

\bibitem{cohen_01}
E.G.D. Cohen, Physica A {\bf 305}, 19 (2002); E.G.D. Cohen,
\emph{Boltzmann and Einstein: Statistics and dynamics - And unsolved  
problem}, Boltzmann Award Communication at Statphys 22 - Bangalore
2004, Pramana (2005), in press. 

\bibitem{casetti_01}
L. Casetti, M. Pettini and E.G.D. Cohen, Phys. Rep. 
{\bf 337}, 237 (2000). 

\bibitem{frenkel}
D. Frenkel and B. Smit, {\it Understanding Molecular Simulation} 
(Academic Press, San Diego, 1996).

\bibitem{latora_01}
V. Latora , A. Rapisarda  and S. Ruffo, 
Phys. Rev. Lett. {\bf 80}, 692  (1998); 
V. Latora, A. Rapisarda and C. Tsallis, Phys. Rev. E 
{\bf 64}, 056134 (2001). 

\bibitem{gell_mann_01}
M. Gell-Mann and C. Tsallis, eds., 
{\it Nonextensive Entropy -- Interdisciplinary Applications}, 
(Oxford University Press, New York, 2004); see also C. Tsallis, M. Gell-Mann, Y. Sato, 
cond-mat/0502274.

\bibitem{murilinho}
de Oliveira, P. M. C., Penna, T. J. P., Herrmann, H. J., Eur. Phys. J. B 
{\bf 1}, 205 (1998)

\bibitem{yoshida_01} 
F. Neri, {\it Lie algebras and canonical integration}, 
Department of Physics, University of Maryland, preprint (1988);
H. Yoshida, Phys. Lett. A {\bf 150}, 262 (1990).

\end{thebibliography}
\end{document}